# Machine learning-enabled high-entropy alloy discovery

Ziyuan Rao[1], PoYen Tung[1,2], Ruiwen Xie[3], Ye Wei[1*], Hongbin Zhang[3], Alberto Ferrari[4], T.P.C. Klaver[4], Fritz Körmann[1,4], Prithiv Thoudden Sukumar[1], Alisson Kwiatkowski da Silva[1], Yao Chen[1,5], Zhiming Li[1,6], Dirk Ponge[1], Jörg Neugebauer[1], Oliver Gutfleisch[1,3], Stefan Bauer[7], Dierk Raabe[1*]

1. Max-Planck-Institut für Eisenforschung GmbH; Düsseldorf, Germany
2. Department of Earth Sciences, University of Cambridge; Cambridge, United Kingdom
3. Institut für Materialwissenschaft, Technische Universität Darmstadt; Darmstadt, Germany
4. Materials Science and Engineering, Delft University of Technology; Delft, The Netherlands
5. School of Civil Engineering, Southeast University; Nanjing, China
6. School of Materials Science and Engineering, Central South University; Changsha, China
7. KTH Royal Institute of Technology; Stockholm, Sweden

*Corresponding authors. Emails: y.wei@mpie.de; d.raabe@mpie.de

**High-entropy alloys are solid solutions of multiple principal elements, capable of reaching composition and feature regimes inaccessible for dilute materials. Discovering those with valuable properties, however, relies on serendipity, as thermodynamic alloy design rules alone often fail in high-dimensional composition spaces. Here, we propose an active-learning strategy to accelerate the design of novel high-entropy Invar alloys in a practically infinite compositional space, based on very sparse data. Our approach works as a closed-loop, integrating machine learning with density-functional theory, thermodynamic calculations, and experiments. After processing and characterizing 17 new alloys (out of millions of possible compositions), we identified 2 high-entropy Invar alloys with extremely low thermal expansion coefficients around $2\times10^{-6}$ K$^{-1}$ at 300 K. Our study thus opens a new pathway for the fast and automated discovery of high-entropy alloys with optimal thermal, magnetic and electrical properties.**

**One-Sentence Summary**: A general approach for high-entropy alloy discovery integrating machine learning with density-functional theory, thermodynamic calculations, and experiments.

Alloy design refers to a knowledge-guided approach for the development of high-performance materials. It was established in the Bronze Age and further developed until today. It is the basis for the discovery of novel materials that enable technological progress. Several thousand metallic alloys have been discovered so far that actually serve in engineering applications. The first essential alloy groups discovered, such as bronze and steel, are all based on one main element which forms the matrix of the material. Over time, alloys with a higher number of alloying elements in larger fractions have been discovered, such as austenitic stainless steels. Today, with the development of high-entropy alloys (HEAs), we have reached a new stage where multiple elements are used in similar fractions (*1, 2*). Considering only the most used elements of the periodic table, this spans a composition space of at least $10^{50}$ alloy variants, a space so large that it is impossible to tackle it by conventional alloy design methods (*3*). These conventional methods for designing alloys, which have been applied to small subspaces of the HEAs composition realm, include the CALculation of PHAse Diagrams (CALPHAD) and Density-Functional Theory (DFT) (*4-6*). However, CALPHAD provides equilibrium-phase diagrams only, while DFT is computationally costly and cannot be readily applied to higher temperatures and disordered alloys (*5, 7*). Likewise, combinatorial experiments (*8*) are very labor-intense and can cover only small regions of the huge composition space of HEAs.

Because of these methodological limitations to find new materials with promising functional and mechanical features, we present here a new approach to accelerate the discovery of HEAs. It is based on the use of machine learning (ML) techniques, with a focus on probabilistic models and artificial neural networks. One conventional ML approach in alloy design is to build a surrogate model that predicts the properties of a given composition, which is then used for candidate inference (*9, 10*). Such a model relies on a set of manually selected (or automatically learned) compositional, structural and atomic features. The advantages of this method include conceptually straightforward implementation and fast inference speed. Its effectiveness and applicability are however limited by the number of available training data and also subject to possible human bias during feature selection. Recently, active learning emerges as an alternative choice for functional materials discovery (*11, 12*). Active learning is a subfield of ML in which surrogate models iteratively select unseen data points that are most informative to improve the predictive power of the models (*13*). The next set of experiments is guided by the previous model trained based upon the results seen so far, yielding new data points that will again be used iteratively for updating the model. Active learning has the potential to reduce the computational costs of alloy design and to both, incorporate and guide experimental data and routines. Although simulation-based active learning methods have shown to yield promising results, the accuracy required for alloy design is challenged by the complicated nature of HEAs (*11, 14*). Yet, purely experimentally based active learning falls short on physical insight and requires multiple copious experimental iterations (*15*).

To overcome these obstacles, we propose here an active learning framework for composition discovery of HEAs, which is efficient for very sparse experimental data. The approach comprises machine learning-based techniques, density-functional theory, mean-field thermodynamic calculations, and experiments. We focus on the design of high-entropy Invar alloys with low thermal expansion coefficient (TEC), for several reasons: i) there is high demand for novel Invar alloys, for serving the huge emerging markets for the transport of liquid hydrogen, ammonia and natural gas; ii) the mechanical properties of the original $Fe_{65}Ni_{35}$ alloy for which Charles Edouard Guillaume received the 1920 physics Nobel price and which is used until today leave room for improvement; iii) alternative Invar alloys (e.g., intermetallic, amorphous or antiferromagnetic Invar compounds) come at forbiddingly high alloy costs and/or poor ductility (*16, 17*); iv) although a few HEAs have the potential to fill this gap (*18, 19*), the lowest TEC ($\sim10\times10^{-6}$ $K^{-1}$) of HEAs reported in the literature exceeds the value of the original $Fe_{65}Ni_{35}$ alloy ($\sim1.6\times10^{-6}$ $K^{-1}$) (*19*). v) our active learning framework mainly considers compositional information instead of the alloy manufacturing process, which makes the Invar effect an ideal target, as these alloys are mostly determined by composition and less by processing (*6, 19*), i.e. it is permissible to use for this task an ML model which ignores microstructure. The background of this latter point is explained in Supplemental Fig. S1 and Table S1.

**Results and discussion**

The active learning framework includes three main steps: latent space sampling for candidate generation, physics-informed screening and experimental feedback (as shown in Fig. 1). The first step towards novel HEA discovery is to generate promising composition candidates. Considering a large number of possible composition combinations of HEAs and the small experimental datasets (see Supplementary Fig. S2), it is challenging to directly sample new compositions with desired properties. Therefore, we developed a HEA COmposition Generating Scheme (HEA-COGS) based on a deep generative model (GM) (*20*). To efficiently generate potential novel functional alloys (e.g., with low TEC), the HEA-COGS leverages unsupervised learning in combination with a stochastic sampling process. Although similar approaches have been widely adopted for chemical compound discovery (*21-23*), HEA-COGS is explicitly optimized for small-to-medium (100-1,000) experimental datasets. The reason for this special purpose ML approach is the fact that – compared to real big data problems – property and composition datasets available in materials science and engineering are comparably sparse.

In the second step, the selected candidates from the HEA-COGS are further processed by a Two-stage Ensemble Regression Model (TERM), to accurately predict the TEC of the new alloys. The TERM includes two ensemble models which are comprised of multilayer perceptron (MLP) and gradient boosting decision tree (GBDT). During the first stage, a high-throughput compositional and atomic feature (*e.g.*, valence electron concentration, atomic radius)-based ensemble model is established to infer the 1,000 possible Invar compositions from the HEA-COGS-selected candidates. In the second stage, with the additional input of physical properties (i.e., magnetization, Curie temperatures and spontaneous volume magnetostriction) calculated by DFT and from thermodynamic databases, the ensemble model is constructed to narrow down the candidates to only 20-30 from the first stage. The important influence of these physical properties on the Invar effect has been shown in a previous study (*6*). In the last step, a ranking-based policy (see methods) is developed to select the most promising compositions, and the top 3 candidates are experimentally measured in the physical properties measurement system (PPMS) and fed back to the database. We repeat the iteration until the discovery of Invar alloys.

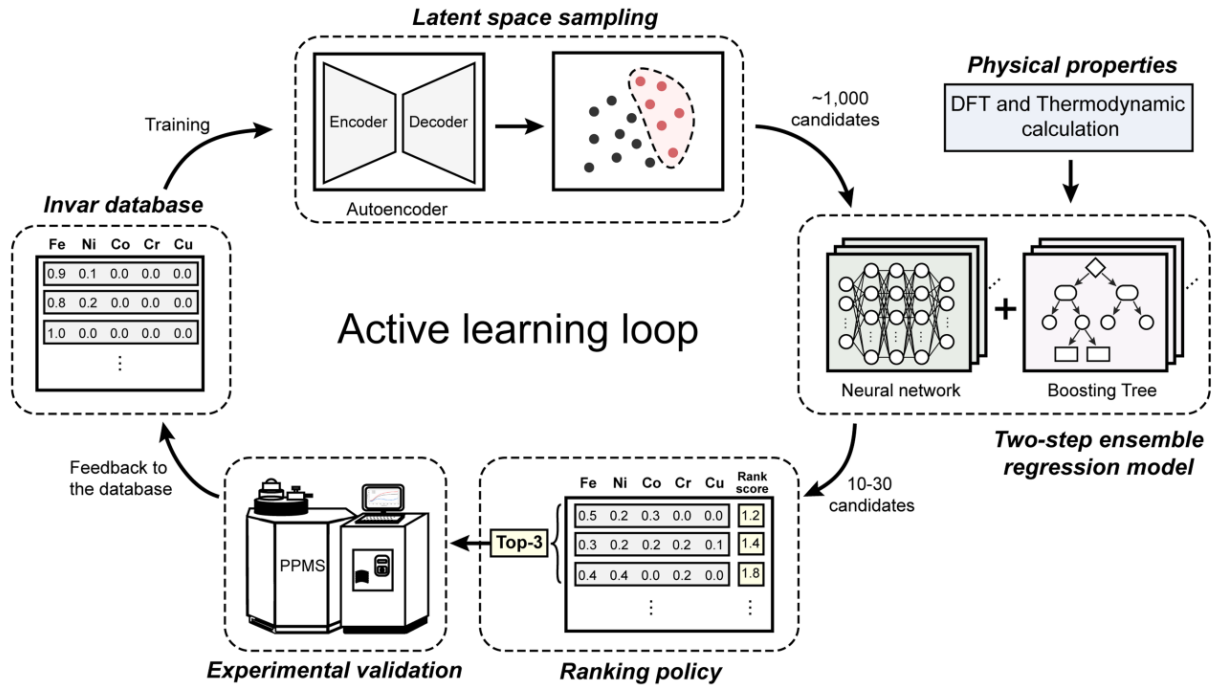

**Figure 1. Approach overview.** An active learning framework for the targeted composition design and discovery of HEAs, which combines machine learning models, DFT calculations, thermodynamic simulations and experimental feedback. In the first step, the promising candidates are generated under the HEA-COGS framework consisting of two primary steps: i) Mixture modeling in low-dimensional latent space. ii) Stochastic sampling for a targeted generation. In the second step, the selected candidates from the HEA-COGS are further processed by TERM framework which includes two ensemble models comprised of multilayer perceptron and gradient boosting decision tree. In the last step, the most promising compositions are selected by a ranking-based policy. The top 3 candidates are experimentally measured and fed back to the database. The iteration is repeated until the discovery of Invar alloys.

In the first step, the HEA-COGS employs a GM to probe the 2-dimensional latent space of alloys (see Fig. 2a). GM has emerged as one of the best models for learning the latent representation of the data (*24, 25*). Different GMs are compared and analyzed based on the evaluation metrics. The results show that the Wasserstein Autoencoder (WAE) Architecture performs better than other models with similar architectures, such as Variational Autoencoder (VAE) (*20, 23*) (see Supplementary Fig. S3). WAE implements a typical two-neural network (NN) architecture, an encoder and a decoder. Different from VAE, WAE attempts to minimize a penalized form of the Wasserstein distance (a mathematical term that measures the distance between two probability distributions) between the encoded data distribution and the target distribution (see Fig. 2a). Here, the encoder $q_\varphi(\mathbf{z}|\mathbf{x})$ with parameters $\varphi$ takes compositions of alloys as the input and learns to compress them down to low-dimensional representations; and the decoder $p_\theta(\mathbf{x}|\mathbf{z})$ with parameters $\theta$ can act as a generator for generating new alloy compositions given the learned continuous latent $z$ representation. Although WAE is trained with only compositional information of alloys, it may implicitly include information of composition-related properties, which makes the latent space physically meaningful and informative. In our case, Invar alloys show extremely low TEC[1] values around room temperature and the composition-TEC relation obeys specific physical laws. One typical example is the effect of VEC on the thermal expansion behavior of 3*d* transitional metals and alloys (see the previous study in (*19*)). Therefore, WAE can also implicitly capture the knowledge of TEC within the latent space of alloys, as shown in Fig. 2a and 2b.

In the second step, the targeted generation (alloy compositions with low TEC) is formulated as a process of sampling from the distribution conditioned on low-TEC $p(\mathbf{x}|\mathbf{c}_{low-TEC})$. With the assumption that the decoder is only related to the given latent representation, the conditional distribution can be expressed a $\int p(\mathbf{x}|\mathbf{z}) p(\mathbf{z}|\mathbf{c}_{low-TEC}) \, d\mathbf{z}$. Using Gaussian Mixture Modeling (GMM) and a property binary classifier (high- or low-TEC) directly trained on the latent $z$ space, we sample the new synthetic compositions with potentially low-TEC from $p(\mathbf{z}|\mathbf{c}_{low-TEC})$ through Markov Chain Monte Carlo (MCMC) sampling (*26*) (See Methods). Introducing the sampling method allows a targeted composition generation from a well-defined probabilistic density distribution instead of a trial-and-error approach or brute-force research in the forward alloy design (technical details are contained in Methods). In the following sections, we demonstrate how HEA-COGS can reduce the computational cost and increase the success rate.

---

[1] The 'TEC' in the following contents means the TEC around room temperature if there is no additional annotation

The TERM is used to further investigate the TEC of the HEA-COGS-generated alloy compositions (in Fig. 2b and 2c). The first stage concerns composition-based regression models, aiming at fast and large-scale composition inference. Then, the top ~1,000 results with potentially low TEC from the first-stage model are screened and enter the second-stage model, where DFT and thermodynamic calculations are included as part of the input, making it a physically informed model. To accelerate the speed and avoid repeated predictions, we avoid calculating the physical properties of all the compositions at one time. Instead, we choose 10-30 compositions with affinity propagation (shown in Supplementary Fig. S4-5) for the DFT and thermodynamic calculations. It is demonstrated that incorporating the physical inputs does significantly increase model accuracy (Supplementary Fig. S6-8). To increase the robustness of TERM without sacrificing the prediction accuracy, TERM taps into the advantages of MLP and GBDT by combining both into a single ensemble. The ensemble models at two stages (in Fig. 2b and 2c) adopts the same basic architecture of TERM. Such architecture consists of 50 MLPs and 50 GBDTs. Theoretically, MLP is capable of approximating any nonlinear relationship, given that the neural layers are sufficiently deep and wide (*27*). While MLP can learn a complex relation, it can also tend to overfit the data if the size of the database is relatively small (*28, 29*). In contrast, GBDT is a state-of-the-art iterative functional gradient algorithm, which has achieved promising results in many modern regression tasks (*30, 31*). GBDT, though sensitive to outliers which can lead to biased predictions (*32*), is a resilient method that curbs overfitting efficiently. Consequently, TERM utilizes the mean TEC prediction value from 100 models (including 50 MLPs and 50 GBDTs) as the final prediction, while the corresponding variance as the uncertainty of the final prediction. Finally, with a ranking strategy (see Methods) the TEC of the top 3 candidates is selected to be experimentally measured by PPMS (in Fig. 2c). The experimental results will augment the training database for the next active learning iteration.

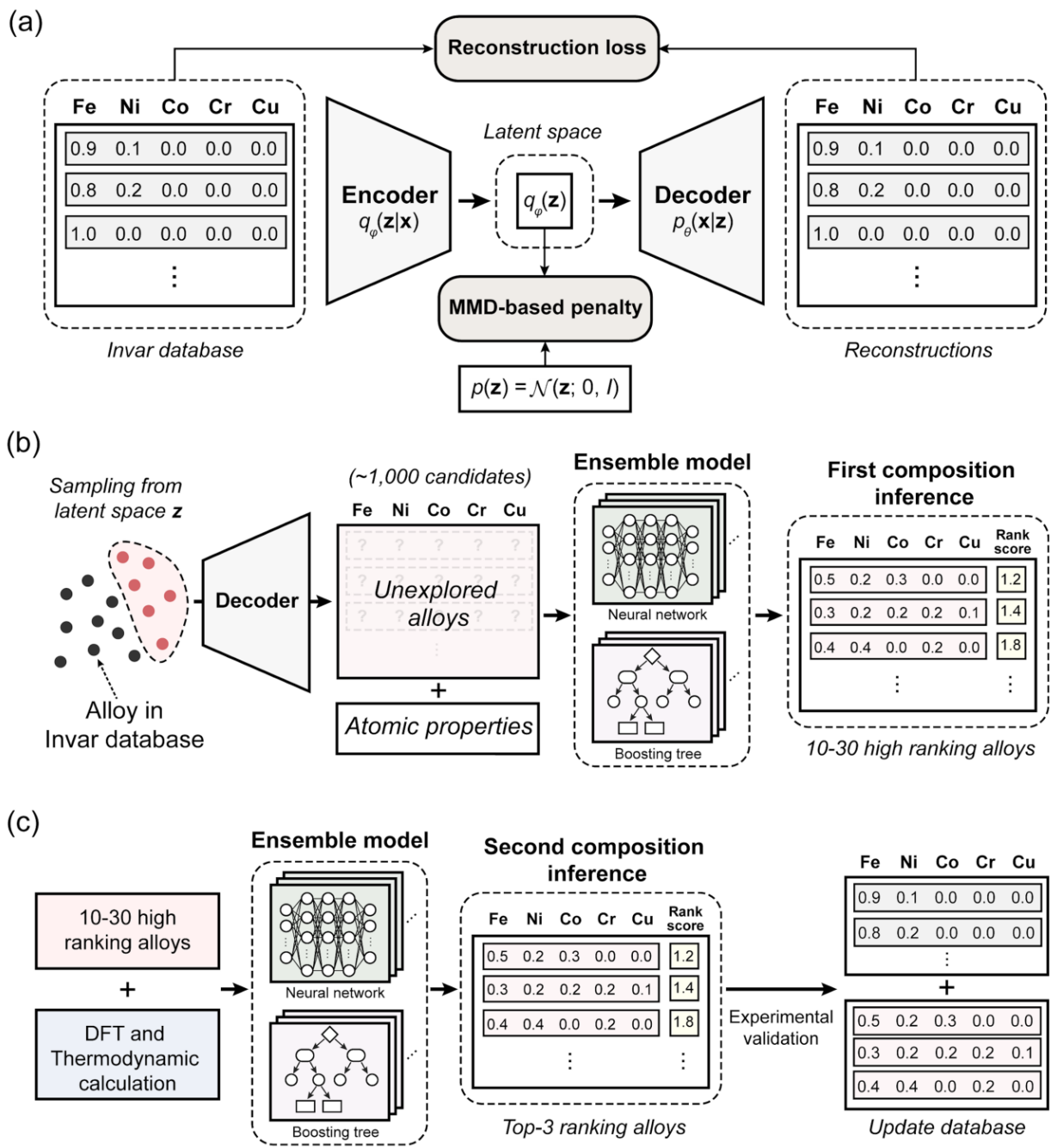

**Figure 2. Schematic framework of the three modeling steps.** (a) A HEA composition generating scheme (HEA-COGS) based on a deep generative model. The encoder $q_\varphi(z|x)$ with parameters $\varphi$ takes compositions of alloys as input and the decoder $p_\theta(x|z)$ with parameters $\theta$ can act as a generator for suggesting new alloy compositions based on the learned latent $z$ representation. (b) A large-scale screening with multilayer neural network model and gradient boosting decision tree.

The new synthetic compositions with potentially low-TEC from $p(z|c_{low-TEC})$ are sampled through the Markov Chain Monte Carlo method based on the latent $z$ space shown in (a). The screening utilizes the mean TEC prediction value from 100 models (including 50 multilayer perceptrons and 50 gradient boosting decision trees) as the final prediction, while the corresponding variance as the uncertainty. A rank-based objective function works as a guide to the final choice. (c) Physically informed screening model with the integration of DFT and thermodynamic calculation. The TEC of the top 3 candidates will be experimentally measured by PPMS and the results will be fed back to the training database for the next active learning iteration.

We produced a large benchmark dataset with 699 data points of Invar alloys mainly from former publications and our unpublished work (see Supplementary Fig. S2 and Table S2). Then based on the HEA-COGS-TERM framework proposed above, we have performed 6 iterations and cast 18 alloys including 17 new alloys and one FeNi classic Invar alloy as a reference alloy. We focus on the design of FeNiCoCr HEAs for the first 3 iterations, and FeNiCoCrCu HEAs for the last 3 iterations. Fig. 3a and 3b show the WAE latent space and GMM-modelled 2D probability density of the first iteration. The latent space yields certain islands that indicate the compositional differences. For example, the HEAs tend to stay in the middle while the binary and ternary alloys tend to stay in the edges of the latent space. Also, a smooth transition among the Fe-Ni, Ni-Co binary alloys and the Fe-Ni-Co ternary alloys can be observed, i.e., the Fe-Ni-Co ternary alloys island is located between the Fe-Ni and Ni-Co binary alloys. FeNiCoCrCu forms a single island, indicating that features of compositions with non-zero Cu content are indeed captured by the HEA-COGS. The new FeCoNiCr HEAs candidates are cross-marked, while the best-ranked HEAs are illustrated with white dots. It is evidenced that the FeCoNiCr candidates are located in the low-TEC region. The results shown above indicate that the learned latent space is informative of TEC of the HEAs.

Fig. 3c and 3d show the last round result of FeCoNiCrCu HEAs discovered by HEA-COGS-TERM (in red color). It is noted that the entire latent space is slightly rotated due to the addition of new data into the training dataset from previous iterations. The augmented dataset also leads to a modified GMM-modelled probability density (in Fig. 3d), in which the left Gaussian eclipse extends more to the left region compared to Fig. 3b. Such phenomena suggest that the HEA-COGS-TERM framework is interpretable and sensitive to the dataset. Considering the same parameters used in the algorithm, even a slight change can lead to a considerably different latent space representation.

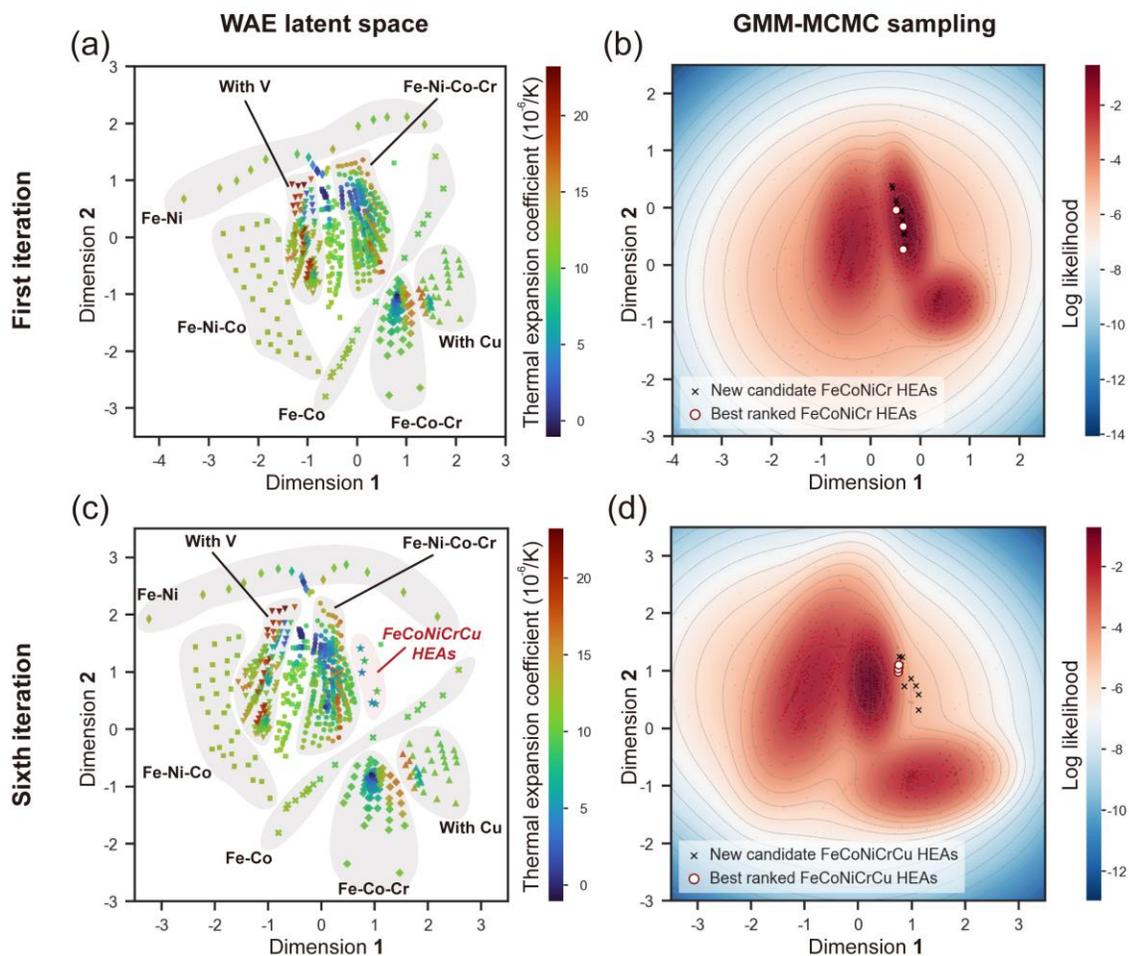

**Figure 3. The first and the last (sixth) iterations of the HEA-COGS generation.** (a) and (b) demonstrate the WAE latent space and Gaussian mixture modeling-modelled density of the first iteration. (c) and (d) demonstrate the WAE latent space and Gaussian Mixture Modeling-modelled density of the last iteration. The WAE latent space distribution of the different compositions is marked with different symbols. The colors of the data points in the latent space denote their corresponding TEC. The Gaussian Mixture Modeling shows the probabilistic density in the latent space. The new candidates proposed by the first stage of the TERM are marked by cross while the new compositions proposed by the second stage of the TERM are marked by circles. The learned latent spaces are informative of TEC of the HEAs.

Fig. 4a provides the concentration histogram of Cr and Cu in the current dataset, respectively. It is noted that the Cr histogram has a long tail where various concentrations (from 0 to 20%) can be found. In contrast, the vast majority (more than 95%) of the compositions have zero or near zero Cu concentration. The lack of experimental data is a common problem encountered in materials science and such distributional difference likely accounts for a significantly different learning behavior observed in Fig. 4b.

Fig. 4b shows the measured and predicted TEC values for FeNiCoCr and FeNiCoCrCu HEAs at each learning iteration. Such plots represent the 'learning curve' of the HEA-COGS-TERM model. For FeNiCoCr HEAs, the learning curve indicates a progressive exploitation endeavor with the learning arrow pointing downwards. This means that the average experimentally measured TEC value gradually decreases: $6.49 \times 10^{-6}$/K in the first, $5.61 \times 10^{-6}$/K in the second and $3.65 \times 10^{-6}$/K in the third iteration (see also Table 1). Interestingly, the model predictions deviate considerably from their experimental counterparts at the beginning, *i.e.*, in the first iteration: the mean and standard deviation of the predicted TEC values are $3.64 \times 10^{-6}$/K and $0.66 \times 10^{-6}$/K, respectively, while the experimentally observed TEC values are $6.49 \times 10^{-6}$/K and $5.03 \times 10^{-6}$/K. This means that the deviation in terms of the mean TEC is 77.5%. Noteworthy, Alloy A3 (see Table 1 below) has the highest predicted TEC value ($4.39 \pm 0.79 \times 10^{-6}$/K), but the experimentally observed TEC shows exactly the opposite, namely, the lowest measured TEC ($1.41 \times 10^{-6}$/K). In the second and third iterations, the standard deviation of the experimentally measured TEC values declines remarkably ($3.34 \times 10^{-6}$/K and $1.46 \times 10^{-6}$/K), and the deviation in terms of the mean also narrows down quickly (from 60.0% to 40.0%). This demonstrates excellent exploration progress in which HEA-COGS-TERM converges quickly and can predict TEC with high accuracy after only three iterations. Just like the human learning process, HEA-COGS-TERM grasps the complex relation of the FeNiCoCr HEAs in a gradual manner. Moreover, it was proposed that there is a law of substitution between the joint concentrations of (Co+Cr) and of Ni (see Supplementary Fig. S9), and indeed we observe that the HEA-COGS-TERM has learned this composition law for Invar alloys by itself and successfully predicted new Invar compositions based on this self-acquired knowledge. On the other hand, a different learning behavior is observed for FeNiCoCrCu. The overall learning tendency shows no significant downward trend of the mean experimentally measured TEC value, from $6.26 \times 10^{-6}$/K for the first iteration, to $6.64 \times 10^{-6}$/K in the second and $5.67 \times 10^{-6}$/K in the third iteration (more numerical details in Table 1). Such a difference can be attributed to the lack of Cu-containing FeNiCoCrCu data (only three data points are available at the beginning, see also Fig. 4a). Despite this shortcoming, the deviation in terms of the mean of the experimentally measured TEC value narrows down, from 33.9% for the first iteration to 10.2% in the last iteration, indicating a gradually improved prediction accuracy. To summarize, we show here that HEA-COGS-TERM is able to discover 7 new Invar alloys with TEC values below $5\times10^{-6}$ $K^{-1}$, from a comparably tiny data array of only 17 cast reference materials. It also identifies the complex relationships between chemical composition and the TEC values. This is an excellent result, given Mosumoto's work that the likelihood to find new Invar alloys is around 0.8% by trial and error (*33-38*). Here, we introduced an approach with more than 50 times higher efficiency compared to the usual trial and error approach.

**Table 1. Compositions and TEC of the HEAs designed in this work.**

| Alloys | Round | Fe (wt. %) | Ni (wt. %) | Co (wt. %) | Cr (wt. %) | Cu (wt. %) | Predicted TEC ($\times10^{-6}$/K) | Predicted uncertainty ($\times10^{-6}$/K) | Experimental TEC ($\times10^{-6}$/K) |
|---|---|---|---|---|---|---|---|---|---|
| A1 | 1st | 0.552 | 0.239 | 0.167 | 0.042 | 0 | 3.41 | 1.29 | 7.54 |
| A2 | 1st | 0.492 | 0.172 | 0.271 | 0.065 | 0 | 3.13 | 0.75 | 10.52 |
| A3 | 1st | 0.418 | 0.094 | 0.409 | 0.080 | 0 | 4.39 | 0.79 | 1.41 |
| A4 | 2nd | 0.525 | 0.220 | 0.208 | 0.047 | 0 | 3.91 | 0.53 | 7.97 |
| A5 | 2nd | 0.440 | 0.138 | 0.346 | 0.076 | 0 | 4.20 | 0.96 | 3.24 |
| **A6** | ***2nd** | **0.635** | **0.365** | **0.000** | **0.000** | **0** | **1.57** | **0.51** | **1.30** |
| A7 | 3rd | 0.424 | 0.126 | 0.377 | 0.073 | 0 | 4.58 | 1.40 | 4.09 |
| A8 | 3rd | 0.442 | 0.158 | 0.332 | 0.068 | 0 | 5.88 | 2.17 | 4.83 |
| A9 | 3rd | 0.541 | 0.228 | 0.172 | 0.059 | 0 | 5.16 | 1.43 | 2.02 |
| B1 | 4th | 0.400 | 0.069 | 0.395 | 0.079 | 0.057 | 7.57 | 1.45 | 5.84 |
| B2 | 4th | 0.488 | 0.178 | 0.222 | 0.062 | 0.050 | 5.48 | 1.01 | 4.38 |
| B3 | 4th | 0.570 | 0.164 | 0.146 | 0.051 | 0.069 | 4.43 | 1.33 | 8.56 |
| B4 | 5th | 0.406 | 0.069 | 0.383 | 0.092 | 0.050 | 8.41 | 1.70 | 4.94 |
| B5 | 5th | 0.577 | 0.229 | 0.083 | 0.052 | 0.059 | 4.50 | 1.00 | 5.31 |
| B6 | 5th | 0.516 | 0.068 | 0.275 | 0.078 | 0.063 | 9.32 | 3.49 | 9.68 |
| B7 | 6th | 0.483 | 0.178 | 0.209 | 0.079 | 0.051 | 5.49 | 0.92 | 5.60 |
| B8 | 6th | 0.500 | 0.183 | 0.183 | 0.080 | 0.054 | 5.65 | 1.16 | 5.13 |
| B9 | 6th | 0.507 | 0.199 | 0.158 | 0.079 | 0.057 | 5.56 | 1.05 | 6.29 |

* This original $Fe_{63.5}Ni_{36.5}$ Invar (A6) is a reference alloy

Fig. 4c shows the TEC as a function of temperatures for 2 Invar alloys (TEC≈$2\times10^{-6}$ $K^{-1}$) discovered in this work (A3 and A9 in Table 1) compared with HEAs and medium entropy alloys (MEAs) (*19, 39*). The alloys discovered in this work show abnormally low TEC compared to HEAs and MEAs reported before. Table 1 shows the compositions and TECs of the 18 alloys experimentally measured in this work. A3 and A9 HEAs with 4 principal elements show extremely low TECs which are comparable with the classical FeNi binary Invar material (A6 in Table 1). B2 and B4 HEAs (in Table 1) with 5 principal elements show TECs 75.6% lower than the prototype reference HEA, i.e., Cantor alloy, and 56.2% lower than the lowest TEC (~$10\times10^{-6}$ $K^{-1}$) in HEAs, i.e. TEC of FeCoNiMnCu HEA, reported so far to the authors' knowledge (*19*). The microstructure and the thermal expansion curves of the alloys are presented in Supplementary Fig. S10-13. The effect of microstructure on the thermal expansion behavior of the Invar alloys is very small (see Supplementary Fig. S1 and Table S1). Fig. 4d shows the configurational entropy against TEC for different kinds of alloys. Most HEAs, MEAs and conventional alloys show high TEC values. On the contrary, most Invar alloys show a low TEC but also low configurational entropy. The Invar alloys discovered in this work show a good combination of low TEC and high configurational entropy. This indicates the high potential of the HEA concept for the design of Invar alloys, which, beyond their beneficial thermal expansion response, also offer high strength, ductility and corrosion resistance.

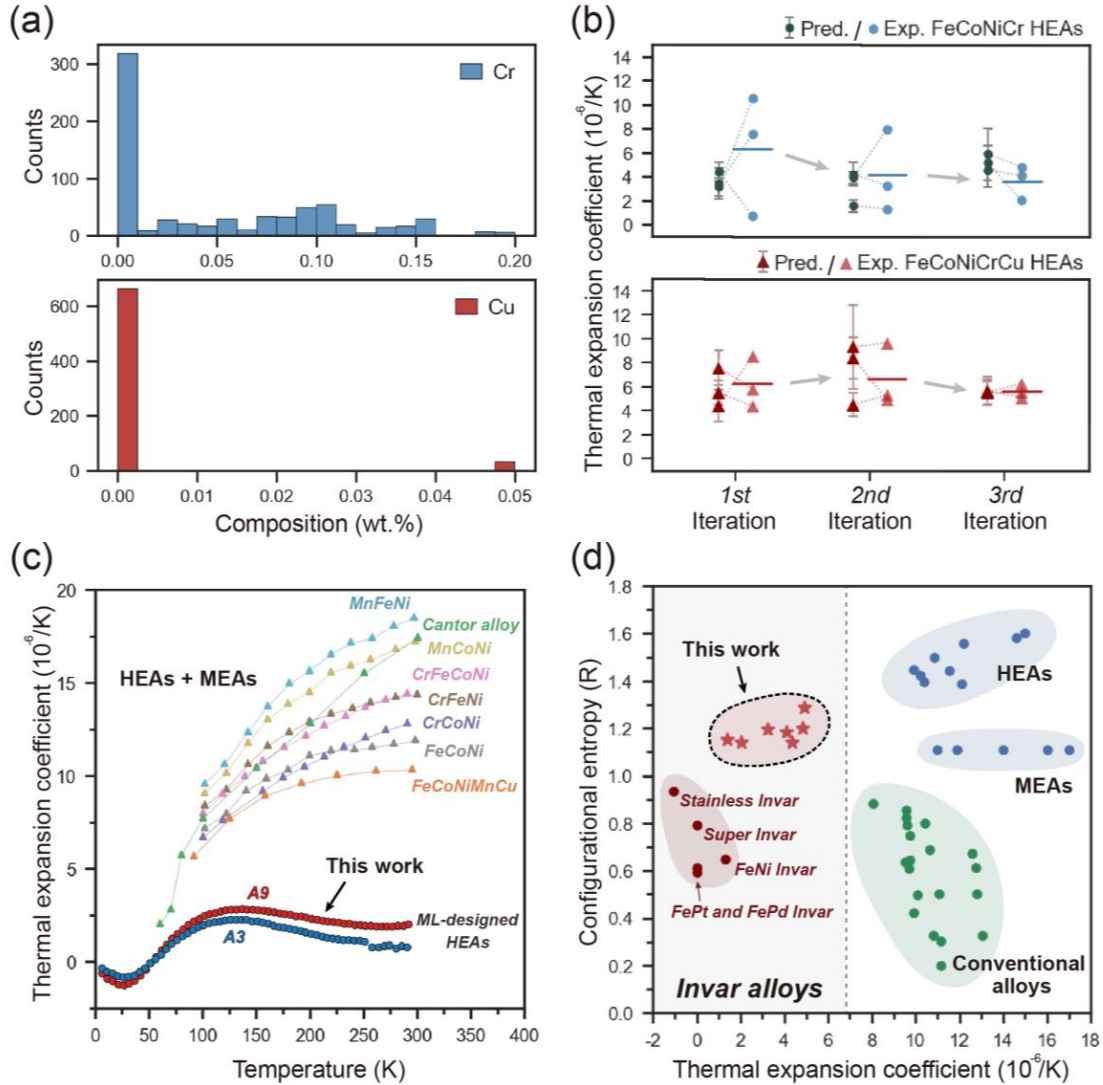

**Figure 4. Summary of the thermal expansion behavior of the new HEAs.** (a) Cr and Cu distributions in all the compositions of the dataset. The Cr histogram has a long tail where various concentrations (from 0 to 20%) can be found. In contrast, the vast majority (more than 95%) of the compositions have zero or near zero Cu concentration. (b) TEC of the FeNiCoCr and FeNiCoCrCu HEAs we cast in all the iterations. Predictions and experimental results are marked in triangles and rectangles respectively. For FeNiCoCr HEAs, the learning curve indicates a progressive exploitation endeavor with the learning arrow pointing downwards. For FeNiCoCrCu HEAs, the deviation in terms of the mean of the experimentally measured TEC value narrows down, indicating a gradually improved prediction accuracy. (c) The thermal expansion behavior of the ML-designed HEAs with low TEC discovered in this work. As a comparison, we also plot

the thermal expansion curve of the HEAs and MEAs. A3 and A9 HEAs discovered in this work show extremely low TECs around $2\times10^{-6}$ K$^{-1}$ at 300 K. (d) Configurational entropy plotted against the TEC values for HEAs, MEAs and conventional alloys and alloys discovered in this work.

**Conclusions**

Understanding the underlying physics behind composition-property relations is the key challenge in alloy design, a task particularly challenging for the case of compositionally complex materials. In principle HEAs with interesting features can hide in a practically infinite and vastly unexplored composition space, a scenario which puts targeted alloy design to its hardest test. We have therefore developed a universally applicable active learning framework that combines a generative model, regression ensemble, physics-driven learning and experiments for the compositional design of HEAs. Our method demonstrates its proficiency in designing high-entropy Invar alloys, based on very spare experimental data. The entire workflow required only a few months, in contrast to a conventional alloy design approach requiring likely years and many more experiments. We expect that more than one properties can be optimized simultaneously using the COGS-TERM framework in the compositional spectrum of HEAs.

**Acknowledgements**: Z.R., R.X., O.G. and H.Z. appreciate the funding by Deutsche Forschungsgemeinschaft (DFG, German Research Foundation) - Project-ID 405553726 – TRR 270. Y.W. appreciates the funding by BiGmax, the Max Planck Society's Research Network on Big-Data-Driven Materials Science and acknowledges the financial support from the ERC-CoG-SHINE-771602. P.T. acknowledges the financial support by Electron and X-ray microscopy Community for structural and chemical Imaging Techniques for Earth materials (EXCITE) - grant number G106564 and International Max Planck Research School for Interface Controlled Materials for Energy Conservation (IMPRS-SurMat). The contributions of Prof. M. Acet, M. Nellessen, M. Adamek and F. Schlüter are also gratefully acknowledged. The Lichtenberg high-performance computer of TU Darmstadt is gratefully acknowledged for providing computational resources for the EMTO calculations in the present work.

**Author contributions**: Z.R., Y.W. and D.R. conceived the idea; Z.R., Y.W., P.T. and S.B. designed and performed the active learning framework of the research; Z.R. performed the experiments of the research; R.X., H.Z., A.F., P.K. and F.K. performed the DFT calculations; P.T.S., A.K.S., and Z.R. performed the thermodynamic calculations; Z.R., Y.W. and P.T. wrote most parts of the manuscript; P.T. produced the final figures; All authors discussed the results and commented on the manuscript.

**Competing interests:** Authors declare that they have no competing interests.

**Data and materials availability:** Requests for data and materials should be sent to the corresponding authors.

**Supplementary materials:**
Methods
Figures S1 to S17
Tables S1 to S15
References (40-56)